\documentclass[preprint, amsmath,amssymb,aps,prapplied,longbibliography]{revtex4-1}
\usepackage{graphicx}
\usepackage{dcolumn}
\usepackage{amsmath}
\usepackage{here}
\usepackage{bm}
\usepackage{color}

\begin{document}

\title{
Voltage-induced precessional switching 
at zero-bias magnetic field
in a conically magnetized free layer 
}

\author{R. Matsumoto}\email{rie-matsumoto@aist.go.jp}
\author{T. Nozaki}
\author{S. Yuasa}
\author{H. Imamura}\email{h-imamura@aist.go.jp}
\affiliation{ 
$^{1}$National Institute of Advanced Industrial Science and Technology (AIST),
Spintronics Research Center, Tsukuba, Ibaraki 305-8568, Japan
}

\date{\today}

\begin{abstract}
  Voltage-induced magnetization dynamics in a conically magnetized free
  layer with an elliptic cylinder shape is theoretically studied on the
  basis of the macrospin model. It is found that
  an application of voltage pulse can induce the precessional switching of magnetization 
  even at zero-bias magnetic field, which is of substantial importance for device applications 
  such as voltage-controlled nonvolatile memory.  Analytical expressions of the
  conditions for precessional switching are derived.
  \end{abstract}


\maketitle

\section{INTRODUCTION}
Voltage control of magnetic anisotropy (VCMA) in
a ferromagnetic metal layer is a promising technology for
the low-power writing in magnetoresistive
random-access memories (MRAMs)
\cite{weisheit_electric_2007, maruyama_large_2009, duan_surface_2008, 
nakamura_giant_2009, tsujikawa_finite_2009, nozaki_voltage-induced_2010,
nozaki_large_2016,
endo_electric-field_2010, shiota_induction_2012, shiota_pulse_2012,
kanai_electric_2012,shiota_evaluation_2016, grezes_ultra-low_2016, shiota_reduction_2017}
compared to the writing with spin-transfer torque (STT)
\cite{slonczewski_current-driven_1996,berger_emission_1996,myers_current-induced_1999}.
In the magnetic tunnel junctions with perpendicular magnetization
(p-MTJs) which have been the mainstream technology for STT-MRAMs,
voltage-driven writing$-$i.e., magnetization switching$-$has been
demonstrated by an application of a bias voltage with an
appropriate pulse width under a bias magnetic field having an in-plane
(IP) component
\cite{kanai_electric_2012, shiota_evaluation_2016, grezes_ultra-low_2016, shiota_reduction_2017}.

For practical applications, writing at zero-bias magnetic field is
desirable to simplify the device structure and reduce the
fabrication cost. It is known that in magnetoresistive devices the shape
anisotropy field can act as a bias magnetic field in some
situations.  The IP shape anisotropy field is obtained by
microfabricating a ferromagnetic film into an elliptic cylinder shape.
In the case of a perpendicularly magnetized free layer, however,
the IP shape anisotropy field cannot move its magnetization 
from the perpendicular direction because the IP shape anisotropy field is
proportional to the IP component of the magnetization.
It is necessary to tilt the magnetization from the perpendicular
direction to perform the precessional switching of the magnetization
around the IP shape anisotropy field.

\begin{figure}[H]
  \includegraphics [width=0.95\columnwidth] {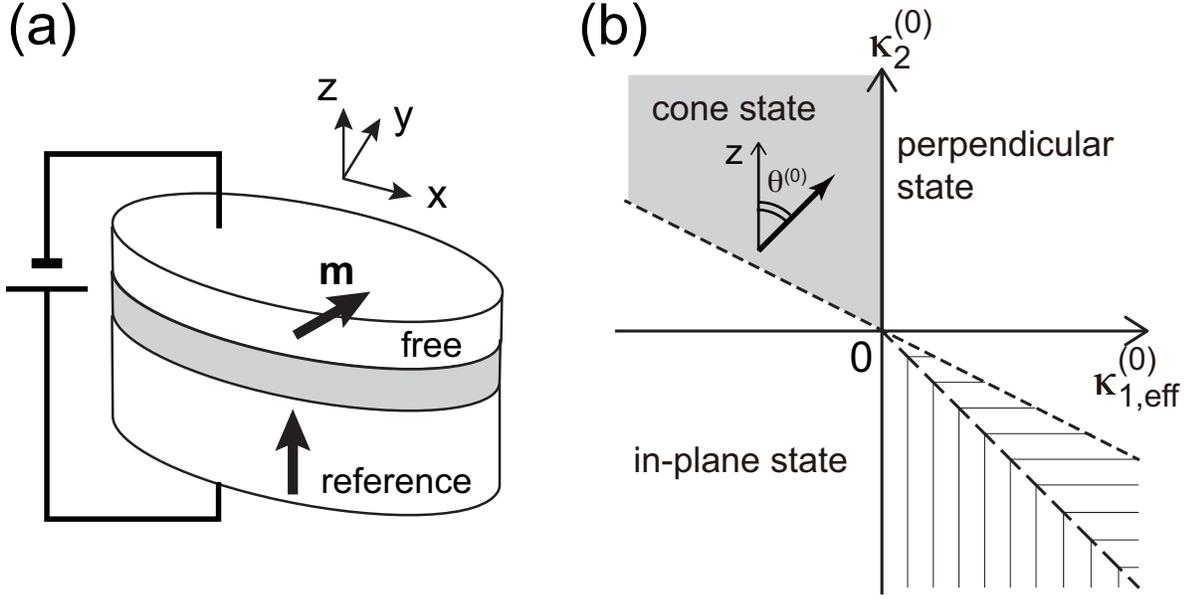}%
  \caption{
  \label{fig:fig1} 
  (a)
  Magnetic tunnel junction with elliptic cylinder shape
  and definitions of Cartesian coordinates $(x, y, z)$.
  The $x$ axis is parallel to the major axis of the ellipse.
  (b) 
  Phase diagram of magnetic film with uniaxial anisotropy
  constants $\kappa_{\rm 1, eff}^{(0)}$ and $\kappa_{\rm 2}^{(0)}$ 
  at equilibrium state (redrawn from
  Ref. \citenum{casimir_rapport_1959}). 
  The shaded area represents the cone-state phase, where the
  film is conically magnetized with equilibrium polar angle, $\theta^{(0)}$. 
  The bistable regions are hatched.
  }
\end{figure}

The titled magnetization state can be obtained by using a conically
magnetized film as a free layer
\cite{apalkov_method_2014, matsumoto_spin-transfer-torque_2015, matsumoto_theoretical_2015}.
The conically magnetized state, which is stabilized by the competition
between the first- and second-order magnetic anisotropy energies, has
been experimentally observed in double-layer and multilayer systems
\cite{lee_spin-reorientation_2002,stillrich_magnetic_2009}.
Recently,  the VCMA effect \cite{park_electric_2016} in addition to the
conically magnetized state
\cite{shaw_perpendicular_2015,fu_temperature_2016,timopheev_second_2016}
has been observed in a Co-Fe-B thin film which is commonly used as a
free layer in an MRAM.

In this paper, the voltage-driven precessional switching in a conically
magnetized free layer with an elliptic cylinder shape is analyzed.
Analytical expressions of the conditions for precessional switching
at zero-bias magnetic field are derived.

\section{MODEL}
The system we consider is schematically shown in Fig. \ref{fig:fig1}(a).
The lateral size of the nanopillar is assumed to be so small that the
magnetization dynamics can be described by the macrospin model. 
The direction of the magnetization in the free layer is
represented by the unit vector ${\bm m} = (m_{x}$, $m_{y}$, $m_{z})
= (\sin \theta \cos \phi$, $\sin \theta \sin \phi$, $\cos \theta$), 
where $\theta$ and $\phi$ are the polar and azimuthal angles of $\bm{m}$.
The magnetization in the reference layer is fixed to align in the
positive $z$ direction.

The energy density of the free layer 
is given by \cite{stiles_spin-transfer_2006}
\begin{align}
  \label{eq:E}
  {\cal E} (m_{x}, m_{y}, m_{z})
  = 
  & 
  \frac{1}{2} \mu_{0} M_{\rm s}^{2}
  ( N_{x} m_{x}^{2} + N_{y} m_{y}^{2} + N_{z} m_{z}^{2} )
  \nonumber\\  
  &
  + K_{\rm u1} (1-m_{z}^{2})  + K_{\rm u2} (1-m_{z}^{2})^{2},
\end{align}
where $K_{\rm u1}$ and $K_{\rm u2}$ are the first- and second-order
anisotropy constants, respectively.
The values of $K_{\rm u1}$ and $K_{\rm u2}$ can be varied 
by application of a bias voltage, $V$, through the VCMA effect. 
The demagnetization coefficients, $N_{x}$, $N_{y}$ and $N_{z}$ are
assumed to satisfy $N_{z} \gg N_{y} > N_{x}$.
$\mu_{0}$ is the vacuum permeability, and 
$M_{\rm s}$ is the saturation magnetization of the free layer.
IP shape anisotropy field, $H_{\rm k}$, is given by  
$H_{\rm k}=M_{\rm s}(N_{y} - N_{x})$ \cite{matsumoto_critical_2016}.
Normalization by $\mu_{0}M_{\rm s}^{2}$ leads to the dimensionless
energy density defined as \cite{stiles_spin-transfer_2006}
\begin{align}
  \label{eq:epsilon}
  {\varepsilon} (m_{x}, m_{y}, m_{z})  =  
  \frac{1}{2}  ( N_{x} m_{x}^{2} + N_{y} m_{y}^{2} + N_{z} m_{z}^{2} )  \nonumber\\
  + \kappa_{\rm 1} (1-m_{z}^{2})  + \kappa_{\rm 2} (1-m_{z}^{2})^{2},
\end{align}
where $\varepsilon={\cal E}/(\mu_{0}M_{\rm s}^{2})$, $\kappa_{\rm
1}=K_{\rm u1}/(\mu_{0}M_{\rm s}^{2})$ and $\kappa_{\rm 2}=K_{\rm u2}/(\mu_{0}M_{\rm s}^{2})$.

Before analyzing the switching conditions, let us show the basic
properties of the equilibrium magnetization state at $V=0$, which is the
initial state of magnetization switching.
The direction of the magnetization of the initial state, ${\bm
m}^{(0)}$, is obtained by minimizing the energy density at $V=0$ as
\cite{casimir_rapport_1959}
\begin{align}
  \label{eq:z0} 
  m_{z}^{(0)} =
  \pm \sqrt{1+ \frac{\kappa_{\rm 1,eff}^{(0)}}{2 \kappa_{\rm 2}^{(0)} }},
\end{align} 
$m_{x}^{(0)}= \pm \sqrt{1- (m_{z}^{(0)})^{2}}$, and $m_{y}^{(0)}=0$. 
Throughout the paper, 
the superscript (0) indicates the
quantities at $V=0$. Without loss of generality, the initial state is
assumed to satisfy $m_{x}^{(0)}>0$ and  $m_{z}^{(0)}>0$. 
The effective first-order anisotropy constant, 
$\kappa_{\rm 1,eff}$, is defined as
$\kappa_{\rm 1,eff} = \kappa_{1} - (1/2) ( N_{z} - N_{x} )$.
Since we are interested in the voltage-induced switching of a conically
magnetized free layer, we concentrate on the cone-state region indicated
by the shaded region in Fig. \ref{fig:fig1}(b), where $\kappa_{\rm 1,eff}^{\rm
(0)}<0$ and $\kappa_{\rm 2}^{(0)}>-(1/2)\kappa_{\rm 1,eff}^{\rm
(0)}$.

The retention time of  an MRAM is 
quantified
by the thermal stability
factor, $\Delta^{(0)}$, which is defined
by the ratio of the energy barrier to the thermal energy, $k_{\rm B} T$,
as
\begin{align}
  \Delta^{(0)}
  =
  \frac{
  \left[ {\cal E}^{(0)} (1, 0, 0)  -
  {\cal E}^{(0)} (m_{x}^{(0)}, 0, m_{z}^{(0)})  \right] V_{\rm F}
  }{
  k_{\rm B} T
  },
\end{align}
where $V_{\rm F}$ represents the volume of the free layer, $k_{\rm B}$
is the Boltzmann constant and $T$ is temperature.
The volume of the free layer is expressed as 
$V_{\rm F}=\pi r_{x}  r_{y} t_{\rm F}$, 
where  $r_{x}$  ($r_{y}$) is
half the length of the major (minor) axis of an ellipse, 
and 
$t_{\rm F}$
is thickness of the free layer.
The thermal stability factor at room temperature should be more than 60
for the retention time over 10 years.

\begin{figure}[H]
\includegraphics [width=0.8\columnwidth] {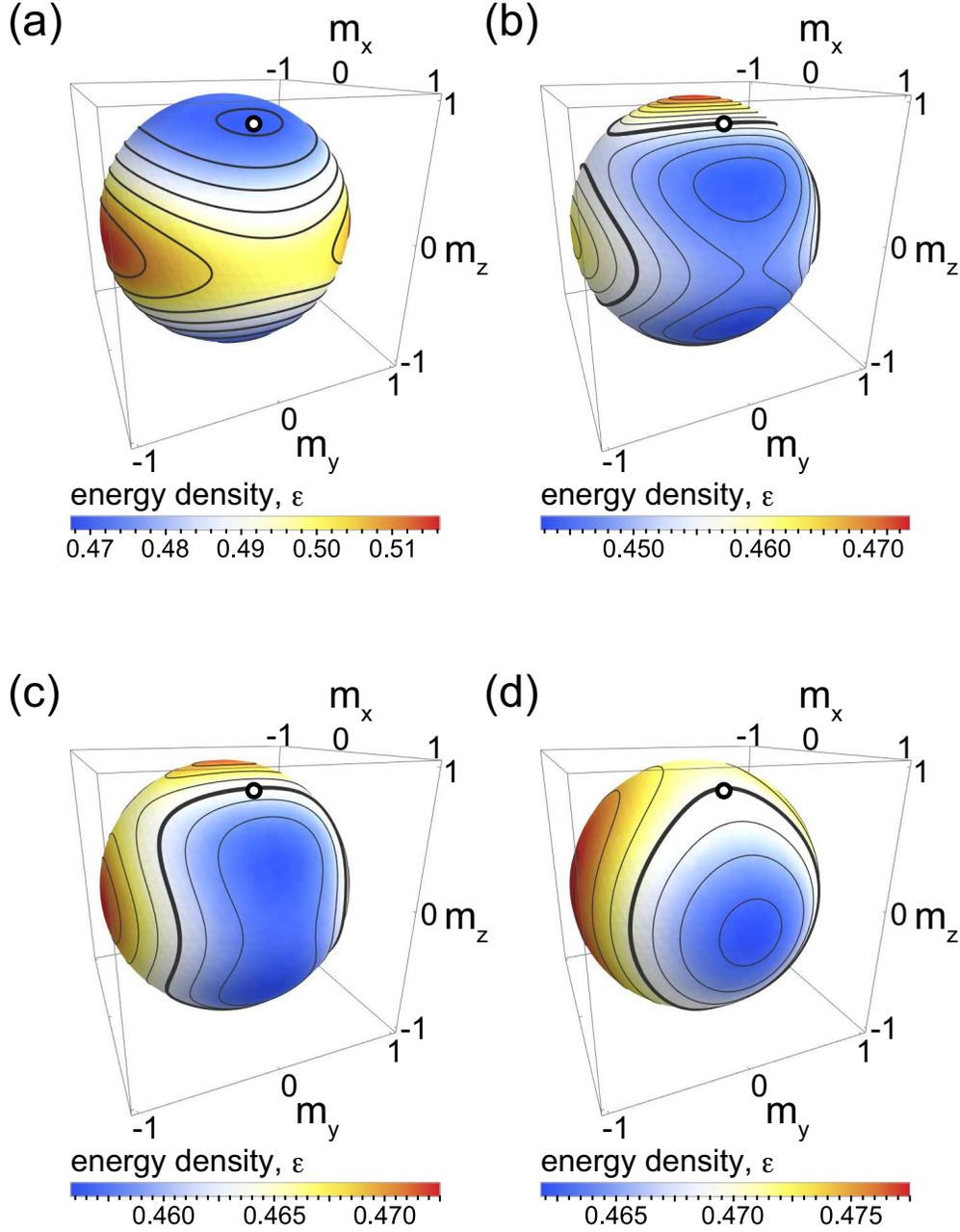}
  \caption{
  \label{fig:eContours} 
  (a) Energy-density contour plot of
  Eq. (\ref{eq:epsilon}) for the initial state, ${\bm m}^{(0)}$.
  The direction of ${\bm m}^{(0)}$ is indicated by the open circle.
  (b) Typical example of the energy-density contour plot of
  Eq. (\ref{eq:epsilon}) in the horizontally hatched region in
  Fig. \ref{fig:fig3}. The anisotropy constants are assumed to be 
  $(\kappa_{1, \rm eff},\kappa_{2}) = ($-0.080, 0.055$)$.
  The open circle indicates ${\bm m}^{(0)}$.
  Thick solid curves represent the contour having the same energy
  density as $\varepsilon(\bm{m}^{(0)})$. 
  (c) The same plot in the vertically hatched region in 
  Fig. \ref{fig:fig3}.
  The anisotropy constants are assumed to be 
  $(\kappa_{1, \rm eff},\kappa_{2}) = ($-0.040, 0.025$)$.
  (d) The same plot in the diagonally hatched region in 
  Fig. \ref{fig:fig3}.
  The anisotropy constants are assumed to be 
  $(\kappa_{1, \rm eff},\kappa_{2}) = ($-0.005, -0.005$)$.
  }
\end{figure}

Although our analysis can be applied to a quite general situation, it
is difficult to understand its benefit without showing specific
examples. The following parameters are assumed:
$M_{\rm s}= 1400$ kA/m, 
$r_{x}=50$ nm, $r_{y}= 20$ nm, 
$t_{\rm F}=1$ nm,
$N_{x}=0.0122$, $N_{y}=0.0443$, $N_{z}=0.9435$
\cite{beleggia_demagnetization_2005},
$H_{\rm k}=566$ Oe,
$K_{\rm 1,eff}^{(0)}=-80$ kJ/m$^{3}$,   
$K_{\rm u1}^{(0)}=1067$ kJ/m$^{3}$,
$K_{\rm u2}^{(0)}=150$ kJ/m$^{3}$,
$m_{z}^{(0)}=0.856$, 
$\Delta^{(0)}$ at $T=300$ K is 61.2.
The equilibrium polar angle is $\theta^{(0)}=31.1^{\circ}$, 
which corresponds to 
a 20\% reduction of
the magnetoresistance ratio from the case of the perpendicularly
magnetized free layer. 
The direction of $\bm{m}^{(0)}$ is indicated by
the open circle in Fig. \ref{fig:eContours}(a) for the contour plot of
$\varepsilon^{(0)}$.

\section{RESULTS}
In voltage-driven precessional switching, only a half period of
precessional motion is used to switch the magnetization. Thanks to the
smallness of the damping constant of the free layer, the trajectory of
the initial half period of precession is well represented by an
energy-density contour including $\bm{m}^{(0)}$ on the Bloch sphere 
(see Appendix A). 
Application of a bias voltage modifies the anisotropy constants,
destabilizes the initial state, and induces the precessional motion.
If the trajectory or energy-density contour starting from $\bm{m}^{(0)}$ crosses
the equator of the Bloch sphere, i.e., $m_{z}=0$, the magnetization can
be switched by turning off the voltage after a half period of 
precession. It is important to find the values of $\kappa_{1,\rm eff}$
and $\kappa_{2}$ which enable precessional switching, i.e., the
conditions for precessional switching.

\begin{figure}[H]
\includegraphics [width=0.95\columnwidth] {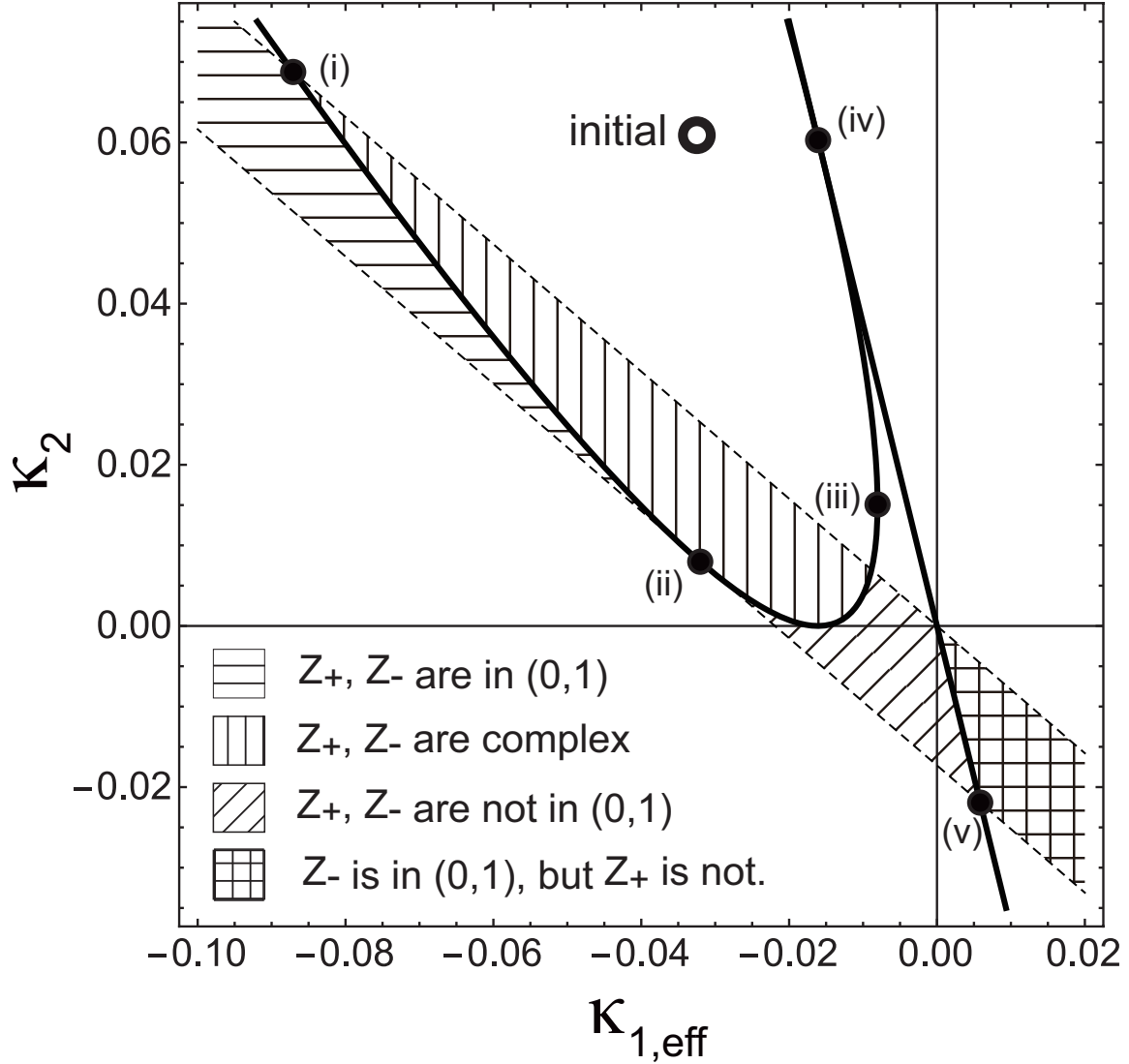}
  \caption{
  \label{fig:fig3} 
  Classification of the values of $Z_{+}$ and $Z_{-}$ in the $\kappa_{1,
  \rm eff}$ and $\kappa_{2}$ plane. The thin dotted lines represent the
  upper and lower boundaries of Eq. \eqref{eq:cond1}. The thick solid
  curves, which are smoothly connected with each other at point (iii),
  represent Eqs. \eqref{eq:bc1} and \eqref{eq:bc2}, respectively.
  The thick solid line represents Eq. \eqref{eq:bc3}.
  In the horizontally hatched region, both $Z_{+}$ and $Z_{-}$ are in the
  region $(0,1)$.  In the vertically hatched region, both $Z_{+}$ and $Z_{-}$
  are complex numbers.  In the diagonally hatched region, both $Z_{+}$
  and $Z_{-}$ are real numbers but not in the region $(0, 1)$.  In the
  crosshatched region, both $Z_{+}$ and $Z_{-}$ are real numbers, but
  only $Z_{-}$ is in the region $(0, 1)$. 
  The point corresponding to the initial state is represented by the open
  circle.  The values of $\kappa_{1, \rm eff}$ at the points indicated
  by the solid circles (i) -- (v) are given in the text.
}
\end{figure}

The energy-density contour having the same energy density as
$\varepsilon(\bm{m}^{(0)})$ is expressed as
\begin{equation}
  \label{eq:traj}
  \varepsilon(\bm{m}) = \varepsilon(\bm{m}^{(0)}).
\end{equation}
The substitution of $m_{z}=0$ into Eq. \eqref{eq:traj} yields
\begin{align}
  \label{eq:ecA1}
  \frac{1}{2}
  \left( N_{x} -N_{y} \right) m_{x}^{2}
  + \frac{1}{2} N_{y} + \kappa_{\rm 1}  + \kappa_{\rm 2}
  = \varepsilon(\bm{m}^{(0)}).
\end{align}
Requiring $0 < m_{x}^{2} < 1$,
one obtains the following inequality
\begin{align} 
  \label{eq:cond1}
  -\xi \kappa_{\rm 1,eff} - \eta
  <
  \kappa_{2}
  <
  -\xi \kappa_{\rm 1,eff},
\end{align}
where
\begin{align}
  \label{eq:xieta}
  \xi = \frac{1}{ 2- (m_{z}^{\rm (0)})^{2}},
  \
  \eta = \frac{\xi(N_{y} - N_{x})}{2(m_{z}^{(0)})^{2}}.
\end{align}
Equation \eqref{eq:cond1} represents the condition for the energy-density
contour to cross the equator. The upper and lower boundaries
of Eq. \eqref{eq:cond1} are indicated by thin dotted lines in
Fig. \ref{fig:fig3}.

There are two kinds of energy-density contour crossing the equator. One
crosses the latitude with $m_{y}=0$ and surrounds the $m_{x}$ axis.
The other crosses the latitude with $m_{x}=0$ and surrounds the $m_{y}$ axis.
The energy density $\varepsilon(\bm{m})$ has symmetry
under a sign change of $m_{x}$. 
Even if $\bm{m}^{(0)}$ is located 
on  the energy-density contour 
surrounding the $m_{y}$ axis,
it connects the initial state $(m_{x}^{(0)}, m_{y}^{(0)}, m_{z}^{(0)})$ with
$(-m_{x}^{(0)}, m_{y}^{(0)}, m_{z}^{(0)})$ and does not result in switching.
The region of $(\kappa_{1, \rm eff},\kappa_{2})$ corresponding
to such trajectories should be excluded in the region defined by
Eq. \eqref{eq:cond1}.

The solutions of Eq. \eqref{eq:ecA1} are given by 
\begin{align}
  \label{eq:ecA2}
  {\tilde m}_{x}
  =
  \pm
  \sqrt{
  \frac{2}{N_{y} -N_{x}}
  \left(
  \frac{1}{2} N_{y} + \kappa_{\rm 1} + \kappa_{\rm 2}
  -\varepsilon(\bm{m}^{(0)})
  \right)
  }.
\end{align}
The energy-density contour crosses the equator, at most, at four points:
$(\pm\tilde{m}_{x},\pm\tilde{m}_{y},0)$, where $\tilde{m}_{y} =
\sqrt{1-\tilde{m}_{x}^{2}}$.  
Therefore, the  energy-density contour 
with $\varepsilon(\bm{m}^{(0)})$ 
corresponding to
the precession around the $m_{x}$ axis does not coexist with that
corresponding to the precession around the $m_{y}$ axis.

The value of $m_{z}$ at which energy-density contour crosses
the latitude with $m_{x}=0$ is obtained as follows. Substitution of
$m_{x}=0$ into Eq. \eqref{eq:traj} leads to
\begin{align} 
  \label{eq:quad}
  &
  \kappa_{2} Z^{2}  
  + \left[
  \frac{1}{2}  ( N_{z} - N_{y} ) -  \kappa_{1}  - 2 \kappa_{2}
  \right]    Z
  \notag\\
  &
  + \frac{1}{2} N_{y} 
  + \kappa_{\rm 1}  + \kappa_{\rm 2} = \varepsilon(\bm{m}^{(0)}),
\end{align}
where $Z=m_{z}^{2}$. The solutions of Eq. \eqref{eq:quad} are obtained as
\begin{align}
  \label{eq:SolZ}
  Z_{\pm}=\frac{ - \left[  \frac{1}{2}  ( N_{z} - N_{y} ) -  \kappa_{1}  - 2 \kappa_{2}  \right]  \pm \sqrt{D}   }{2  \kappa_{2} },
\end{align}
where
\begin{align}
  \label{eq:D}
  D
  &
  =
  \left[
  \frac{1}{2} ( N_{z} - N_{y} ) -  \kappa_{1}  - 2 \kappa_{2}
  \right]^{2} 
  \notag\\
  &
  - 4 \kappa_{2}
  \left[
  \frac{1}{2} N_{y}  + \kappa_{\rm 1}  + \kappa_{\rm 2} -
  \varepsilon(\bm{m}^{(0)})
  \right].
\end{align}
The energy-density contour crosses the latitude with $m_{x}=0$
at $m_{z}=\pm\sqrt{Z_{\pm}}$ if $0 < Z_{\pm} < 1$.

In the region defined by Eq. \eqref{eq:cond1}, the solutions are
classified into four groups as shown in Fig. \ref{fig:fig3}.
In the horizontally hatched region both $Z_{+}$ and $Z_{-}$ are in (0, 1), which means that the energy-density contour crossing the equator
also crosses the latitude with $m_{x}=0$. No energy-density contour
corresponding to the precessional switching exists in this region as shown
in Fig. \ref{fig:eContours}(b).

In the vertically hatched region, the
solutions $Z_{\pm}$ are complex, which means that the energy-density
contour crossing the equator does not cross the latitude with $m_{x}=0$
but surrounds the $m_{x}$ axis. Precessional switching is available,
as shown in Fig. \ref{fig:eContours}(c).
In Fig. \ref{fig:eContours}(c), the anisotropy constants under a bias voltage are assumed to be 
$(\kappa_{1, \rm eff},\kappa_{2}) = ($-0.040, 0.025$)$
which is
$(K_{1, \rm eff}, K_{\rm u2}) = ($-98.5, 61.6$)$ kJ/m$^{3}$ 
in SI units.

The anisotropy constants, for example, can be obtained by
the application of  
$V= 1$ V, i.e., 
the electric field 
($V/t_{\rm I}$)
when the VCMA effect is 
$\eta_{1}=18.5$ fJ/(V m) and $\eta_{2}=88.4$ fJ/(V m).
Here, 
$t_{\rm I}$ represents the thickness of the insulator layer sandwiched between the free and the reference layers,
and $t_{\rm I}=1$ nm is assumed.
the linear bias-voltage dependence of $K_{1, \rm eff}$ and $K_{\rm u2}$ are assumed, and
$\eta_{1}$ and $\eta_{2}$ represent the coefficient of the VCMA effect for $K_{1, \rm eff}$ and $K_{\rm u2}$.
The anisotropy constants per unit area, 
$K_{\rm 1,eff}  t_{\rm F} $ and $K_{\rm u2}  t_{\rm F} $, are expressed as
$K_{\rm 1,eff}  t_{\rm F} = K_{\rm 1,eff}^{\rm (0)}  t_{\rm F}  - \eta_{1} (V/t_{\rm I})$ and
$K_{\rm u2}  t_{\rm F} = K_{\rm u2}^{\rm (0)}  t_{\rm F} - \eta_{2}(V/t_{\rm I})$. 

It should be noted that 
the precessional switching is available
even if $K_{\rm u2}$ is not changed by the bias voltage
\cite{kanai_electric_2014, okada_electric-field_2014}, 
i.e.,  $\eta_{2}=0$.
For example,  
$(\kappa_{1, \rm eff},\kappa_{2}) = ($-0.080, 0.061$)$
is included in the vertically hatched region
while ($\kappa_{1, \rm eff}^{(0)}$, $\kappa_{2}^{(0)})=($-0.032, 0.061$)$ in the initial state.
In SI units, 
$(K_{1, \rm eff}, K_{\rm u2}) = ($-197, 150$)$ kJ/m$^{3}$, and 
it can be obtained 
at $V= 1$ V and $t_{\rm I}=1$ nm, i.e., 
the electric field 
($V/t_{\rm I}$)
of 1 V/nm 
and $\eta_{1}=117$ fJ/(V m). 
Such linear bias-voltage dependence with $\eta_{1} \sim 100$ fJ/(V m)
has been experimentally demonstrated 
for the VCMA effect due to the modulation of charge accumulation without charge trapping
\cite{nozaki_magnetization_2014, skowronski_perpendicular_2015, li_enhancement_2017}.

In the diagonally hatched
region, the solutions $Z_{\pm}$ are real but outside of (0, 1), which means that the energy-density contour crossing the equator
does not cross the latitude with $m_{x}=0$, and precessional
switching is available, as shown in Fig. \ref{fig:eContours}(d).

The boundary among the horizontally hatched, vertically hatched, and
diagonally hatched regions is given by 
\begin{align}
\label{eq:bc1}
  \kappa_{2}
  =
  \frac{- \kappa_{\rm 1,eff} - \sqrt{\Lambda}  }{2(1- Z^{\rm (0)})},
\end{align}
where $Z^{\rm (0)} = (m_{z}^{(0)})^2$ and 
\begin{equation}
  \Lambda
    =
    - \kappa_{\rm 1,eff}  (N_{y} -N_{x})
    -  \frac{1}{4} (N_{y} -N_{x})^{2}.
\end{equation}
This boundary is plotted by the solid curve in Fig. \ref{fig:fig3}.
The curve of Eq. \eqref{eq:bc1} crosses the upper boundary of Eq. \eqref{eq:cond1} at point (i), where
\begin{equation}
\kappa_{1, \rm eff}^{(\rm i)}
  =
  -
  \frac{
  (N_{y} -N_{x})
  (2-Z^{(0)})
  \left(2-Z^{(0)} +2 \sqrt{1- Z^{(0)}} \right)
  }{
  2(Z^{(0)})^{2} }.
\end{equation}
The curve of Eq. \eqref{eq:bc1} is tangential to the lower
boundary of Eq. \eqref{eq:cond1} at point (ii), where
\begin{equation}
  \kappa_{1, \rm eff}^{(\rm ii)}
    =
    - \frac{
    (N_{y} -N_{x})
    \left[
     2 -2 Z^{\rm (0)} + (Z^{\rm (0)})^{2}
    \right]
    }{
    2 (Z^{\rm (0)})^{2} }.
\end{equation}
Equation \eqref{eq:bc1} is obtained by solving  $D=0$,
which has another solution
\begin{align}
  \label{eq:bc2}
  \kappa_{2}
  =
  \frac{- \kappa_{\rm 1,eff} + \sqrt{\Lambda}  }{2(1- Z^{\rm (0)})}.
\end{align}
The curve representing Eq. \eqref{eq:bc2} smoothly connects with that
of Eq. \eqref{eq:bc1} at point (iii), where
\begin{equation}
  \kappa_{1,\rm eff }^{(\rm iii)}
    =-\frac{1}{4} (N_{y} -N_{x}).
\end{equation}

\begin{figure}[H]
\includegraphics [width=0.95\columnwidth] {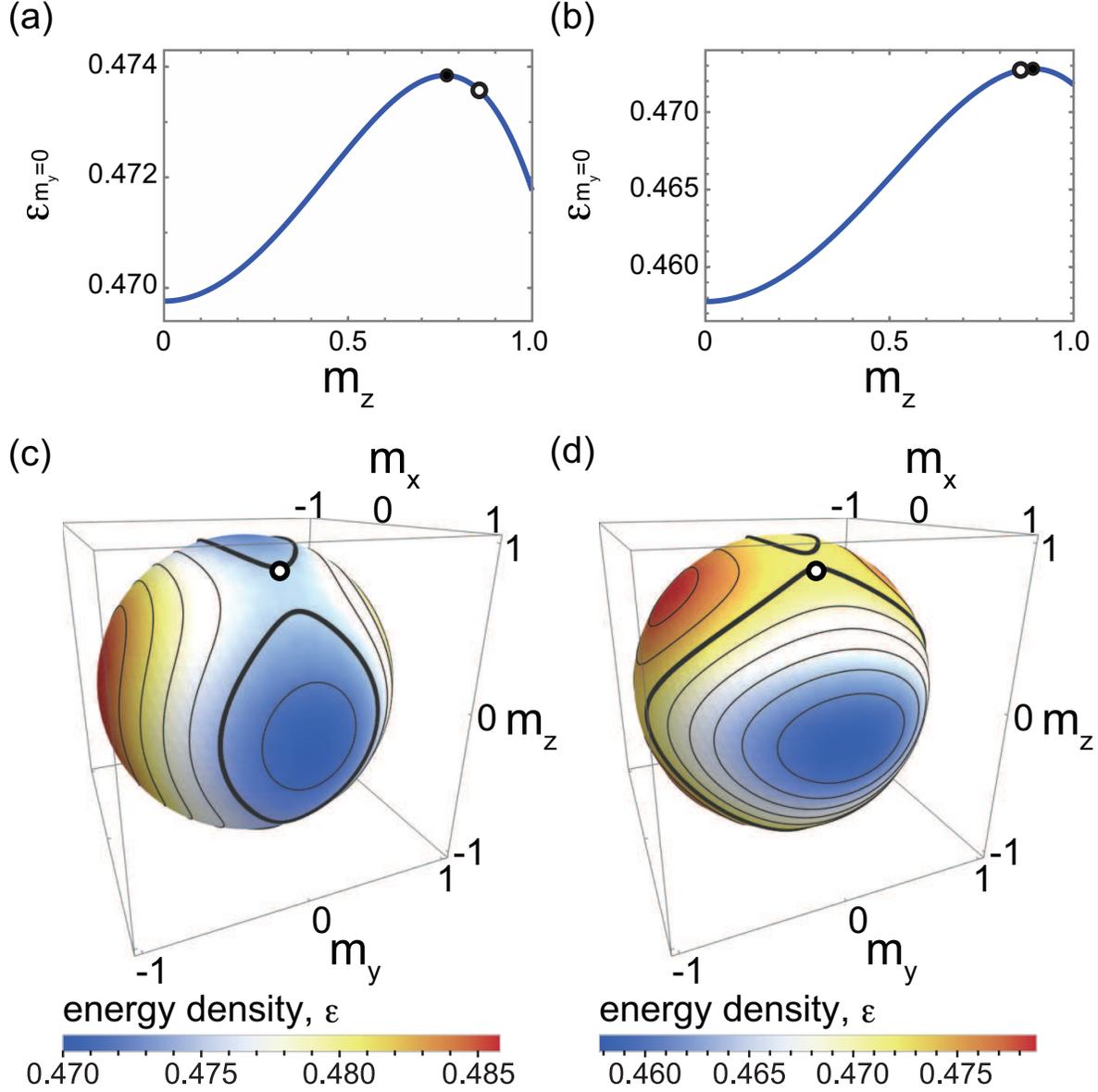}
\caption{\label{fig:fig4} 
  (a) $m_{z}$ dependence of energy density with $m_{y}=0$. The
  anisotropy constants are ($\kappa_{\rm 1,eff}$, $\kappa_{2}$)=(0.010,
  -0.012).  The open and solid circles indicate the initial state, $\varepsilon_{m_{y}=0}(m_{z}^{(0)})$,  
  and the maximum point, respectively.
  (b) The same plot as (a) for the anisotropy constants of
  ($\kappa_{\rm 1,eff}$, $\kappa_{2}$)=(0.010, -0.024).
  (c) Energy-density contour plot for the same parameters as in (a).
  The open circle indicates ${\bm m}^{(0)}$. Thick solid curves represent
  the contour having the same energy density as
  $\varepsilon(\bm{m}^{(0)})$.
  (d) Energy-density contour plot for the same parameters as in (b). 
}  
\end{figure}

In the crosshatched region, only $Z_{-}$ is in (0, 1), and two kinds of energy-density contours coexist. One crosses the
latitude with $m_{x}=0$ at $m_{z} = \pm\sqrt{Z_{-}}$ and corresponds to the
precession around the $m_{z}$ axis. The other corresponds to the
precession around the $m_{x}$ axis.

The boundary between the diagonally hatched region and the cross
hatched region is given by
\begin{align}
  \label{eq:bc3}
  \kappa_{2}
  =
  -\frac{\kappa_{1,\rm eff}}{1-Z^{(0)}},
\end{align}
which is represented by the thick solid line in Fig. \ref{fig:fig3}.
This line is tangential to the curve of Eq. \eqref{eq:bc2} at point
(iv), where 
\begin{equation}
  \kappa_{1,\rm eff }^{(\rm iv)}
    =- \frac{1}{2} (N_{y} -N_{x}).
\end{equation}
Equation \eqref{eq:bc3} crosses the lower boundary of Eq. \eqref{eq:cond1} 
at point (v), where
\begin{align}
  \kappa_{1,\rm eff }^{(\rm v)}
    =
    \frac{
    (N_{y} -N_{x})
    \left(1 -Z^{\rm (0)} \right)
    }{
    2 Z^{\rm (0)}}.
\end{align}
 
In the crosshatched region, the precessional switching is available if
the initial state is located on the energy-density contour surrounding
the $m_{x}$ axis. The condition for the precessional switching is obtained
by analyzing the $m_{z}$ dependence of the energy density on 
the latitude with $m_{y}=0$, $\varepsilon_{m_{y}=0}(m_{z})$.
A typical example of $\varepsilon_{m_{y}=0}(m_{z})$ for parameters with which
the magnetization does not switch is shown in Fig. \ref{fig:fig4}(a).
The anisotropy constants are assumed to be ($\kappa_{\rm 1,eff}$,
$\kappa_{2}$)=(0.010, -0.012).
The initial state,  $\varepsilon_{m_{y}=0}(m_{z}^{(0)})$, is indicated by the open circle and the maximum point
of $\varepsilon_{m_{y}=0}(m_{z})$ is indicated by the solid circle.
Since the $m_{z}$ of the initial state is larger than that of the
maximum point, the initial state is located on the energy-density contour
surrounding the $m_{z}$ axis as shown in Fig. \ref{fig:fig4}(c) and
the magnetization does not switch.

\begin{figure}[H]
\includegraphics [width=0.95\columnwidth] {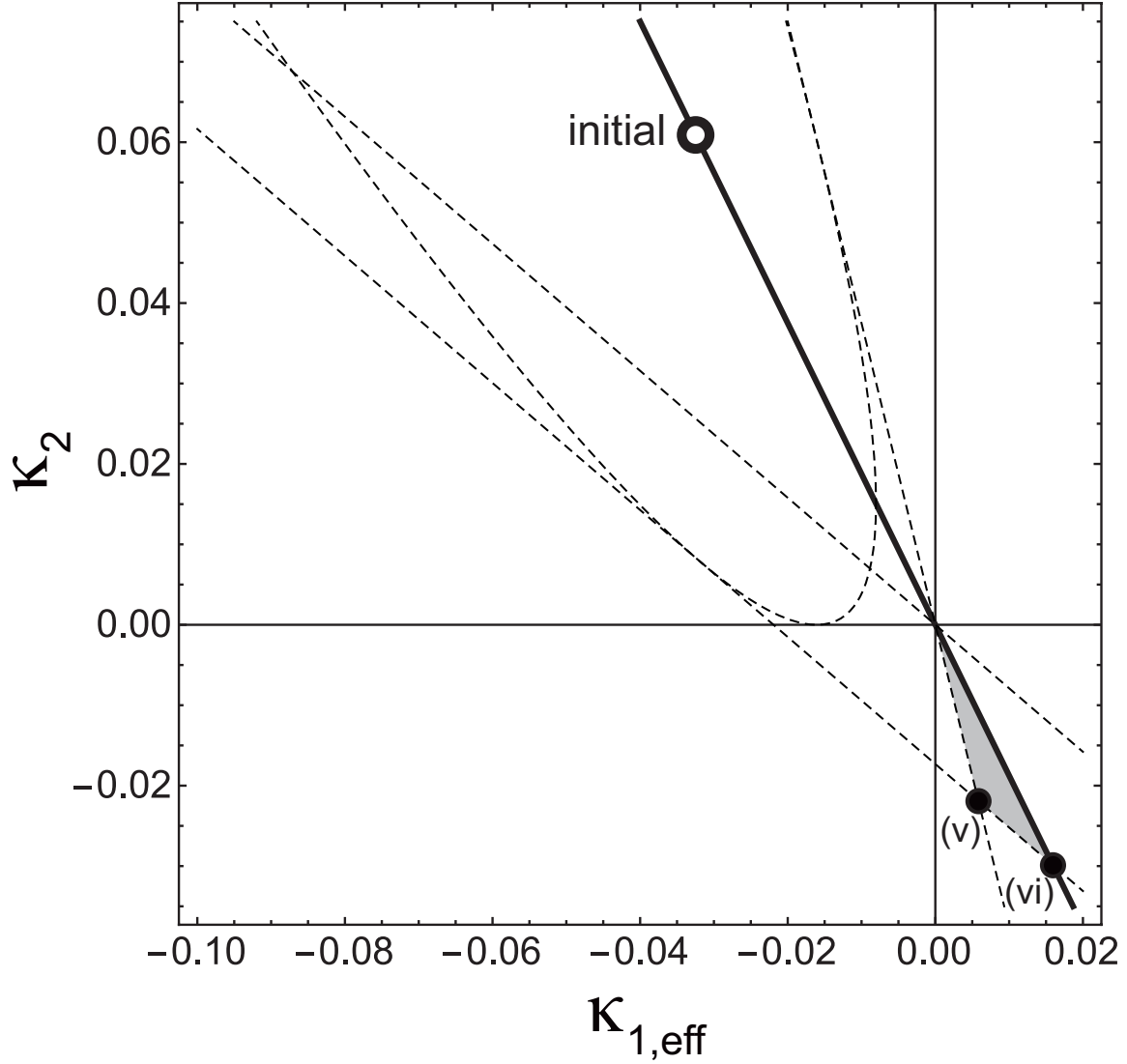}
\caption{\label{fig:fig5} 
  The switching region in the crosshatched area in Fig. \ref{fig:fig3}
  is indicated by the shade triangle with vertices at the origin and 
  at points (v) and (vi). The thin dotted lines and curves represents the
  boundaries shown in Fig. \ref{fig:fig3}.
  The thick solid line represents Eq. \eqref{eq:bc4}.  The point
  corresponding to the initial state is indicated by the open circle.
  The value of $\kappa_{1, \rm eff}$ at the point indicated by solid
  circle (vi) are given in the text.
}  
\end{figure}

A typical example of $\varepsilon_{m_{y}=0}(m_{z})$ for parameters with which
precessional switching is available is shown 
in Fig. \ref{fig:fig4}(b), 
where the anisotropy constants are assumed to be ($\kappa_{\rm
1,eff}$,$\kappa_{2}$)=(0.010, -0.024).
Since the $m_{z}$ value of the initial state is smaller than that at the
maximum point, the initial state is located on the energy-density contour
surrounding the $m_{x}$ axis as shown in Fig. \ref{fig:fig4}(d), and
precessional switching is available.

The boundary of the switching region is obtained by
locating the initial state at the maximum point as
\begin{align}
\label{eq:bc4}
  \kappa_{2} = \frac{\kappa_{2}^{(0)}}{\kappa_{1, \rm
  eff}^{(0)}}\kappa_{1, \rm eff},
\end{align}
which is indicated by the thick solid line in Fig. \ref{fig:fig5}.
The switching region is inside the triangle with vertices at points
(v) and (vi) and the origin as shown by the shaded region in Fig. \ref{fig:fig5}.
The value of $\kappa_{1, \rm eff}$ at point (vi) is given by
\begin{align}
  \kappa_{1, \rm eff}^{(\rm vi)}
  =
  \frac{(N_{y} -N_{x}) (1- Z^{(0)})}{(Z^{\rm (0)})^{2}  }.
\end{align}

\begin{figure}[H]
\includegraphics [width=0.95\columnwidth] {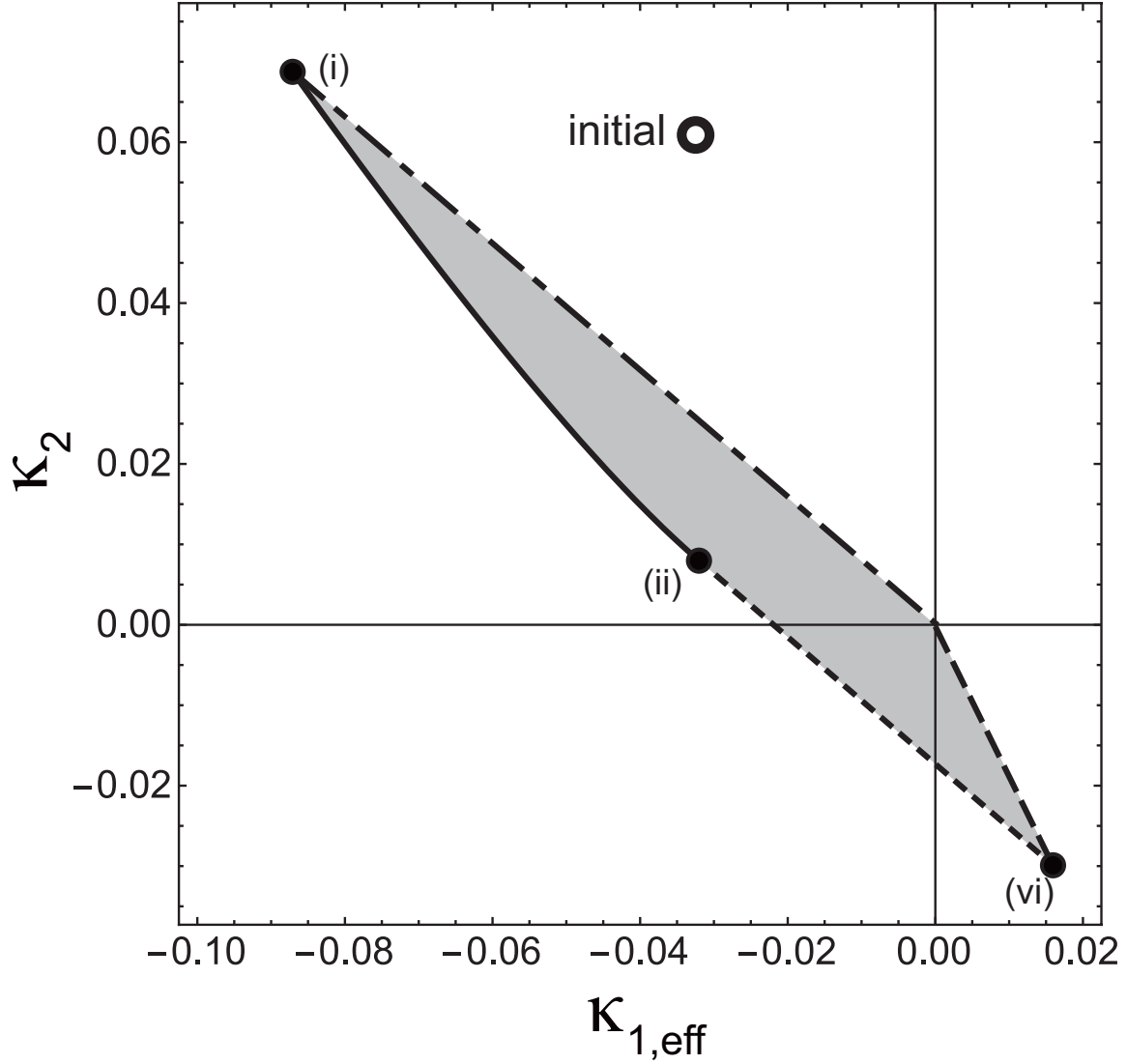}
\caption{\label{fig:fig6} 
 The total switching region and its boundary. The initial values are
 indicated by the open circle. The solid circles represent the
 vertices of the total switching region other than the origin. The
 labels of the vertices are the same as those in Figs. \ref{fig:fig3}
 and \ref{fig:fig5}. The analytical expressions of 
 the boundaries indicated by the solid curve, the dotted line, the dashed line
  and the dotted-dashed line are given in the text.
}  
\end{figure}

Figure \ref{fig:fig6} shows the total switching region, which is given
by the combination of the vertically hatched region, the diagonally hatched
region in Fig. \ref{fig:fig3} and the shaded region in
Fig. \ref{fig:fig5}. The boundary indicated by the solid curve is given
by Eq. \eqref{eq:bc1}. The boundaries indicated by the dotted and
dotted-dashed lines are the lower and upper boundaries of
Eq. \eqref{eq:cond1}, respectively. The boundary indicated by the dashed
line is given by Eq. \eqref{eq:bc4}.

It should be noted that as long as $N_{z} \gg N_{y} > N_{x}$ and
$0 < \theta^{(0)} < \pi/2$, the values
$\kappa_{1, \rm eff}$ at points (i), (ii), and (vi)  satisfy 
\begin{align}
    \label{eq:Points}
  \kappa_{1, \rm eff}^{\rm (i)} <
  \kappa_{1, \rm eff}^{\rm (ii)} <
 0 < 
 \kappa_{1, \rm eff}^{\rm (vi)},
  \end{align}
and the derived analytical expressions for the boundaries of
total switching region are  therefore valid.


Let us make some brief comments on the effects of the
pulse width and finite temperature on switching. 
The pulse width which enables the
precessional switching ranges from about the second quarter to the
third quarter of the precession period. After turning off the voltage
pulse, the magnetization relaxes to the equilibrium state with precessing
around the effective field.
At finite temperature, the initial state distributes around the
equilibrium direction. The thermal distribution of the initial state
is one of the main causes of the write error rate
(WER). The WER of the voltage-controlled MRAM was studied in
Refs. [\citenum{shiota_evaluation_2016}, \citenum{shiota_reduction_2017}] and is known to take a minimum value
at half of the precession period. Similar dependence of the WER on
the pulse width is expected in our system.


\section{conclusion}
In conclusion, voltage-induced magnetization dynamics in a conically
magnetized free layer with an elliptic cylinder shape is studied
theoretically in this paper. It is shown that precessional switching of
magnetization can be performed by applying a voltage pulse even at zero-bias magnetic field. 
The analytical expressions of the conditions for
precessional switching are derived, which is valid as long as the 
conically magnetized free layer is microfabricated into an elliptic
cylinder shape.
The results provide a practical guide for designing a bias-field-free
voltage-controlled MRAM, which simplifies the device structure and
reduces the fabrication cost. 

\acknowledgements
This work was partly
supported by the ImPACT Program of the Council for
Science, Technology and Innovation, and JSPS KAKENHI Grant No. JP16K17509.

\appendix
\section{TRAJECTORY OF PRECESSIONAL SWITCHING}
\label{sec:appendixA}
Our analysis is based on the assumption that the trajectory of
precessional switching is well represented by an energy-density
contour including $\bm{m}^{(0)}$ on the Bloch sphere. In order to show
the validity of this assumption, we perform numerical simulations to
calculate the exact trajectory by solving the Landau-Lifshitz-Gilbert (LLG)
equation \cite{stiles_spin-transfer_2006},
\begin{equation}
  \label{eq:LLG}
    \frac{d{\bm m}}{dt}
    = -\gamma_{0} {\bm m}\times
    {\bm H}_{\rm eff}
    +\alpha
    {\bm m}\times
    \frac{d{\bm m}}{dt},
\end{equation}
where $t$ is the time, $\gamma_{0}$ is the gyromagnetic ratio, and $\alpha$ is the
Gilbert damping constant.
The effective magnetic field, ${\bm H}_{\rm eff}$, 
is defined as
\begin{align} 
  {\bm H}_{\rm eff}= -\frac {1}{\mu_{0} M_{\rm s}} \nabla {\cal E}.
\end{align}

\begin{figure}
\includegraphics[width=0.95\columnwidth]{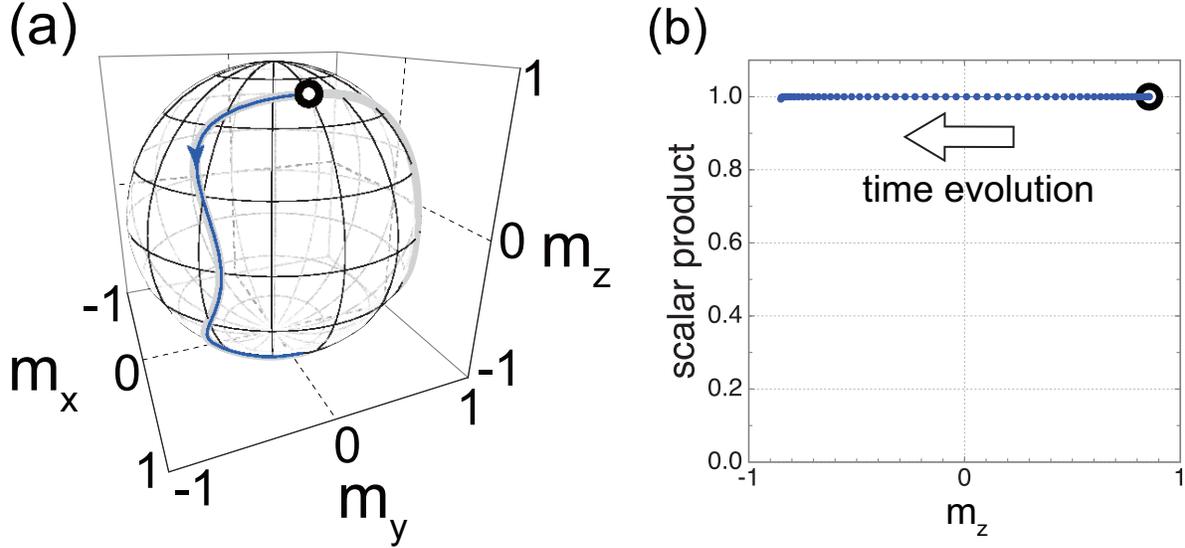}
\caption{\label{fig:fig7} 
  Time evolution of ${\bm m}$ under voltage during precessional switching.
  The parameters are the same as in Fig. \ref{fig:eContours}(c). 
  (a) The simulated trajectory of ${\bm m}$ = ($m_{x}$, $m_{y}$, $m_{z}$)
  is plotted by the thin blue curve together with the corresponding
  energy-density contour (thick gray curve). The direction of the precession is
  indicated by the arrow. The open circle indicates ${\bm m}^{(0)}$.
  (b) The scalar product between magnetization unit vectors of the simulated
  trajectory and the corresponding energy-density contour is plotted as a
  function of $m_{z}$. The direction of the time evolution is indicated by
  the arrow.
}
\end{figure}

Figure \ref{fig:fig7}(a) shows the simulated trajectory (the blue curve)
together with the corresponding energy-density contour (the gray curve)
on the Bloch sphere. The open circle indicates the initial state, ${\bm m}^{(0)}$. The parameters are the same as 
in Fig. \ref{fig:eContours}(c) and the damping constant is assumed to be $\alpha=0.005$.
One can see that within the initial half period of precession, $0 \leq t
\leq 0.56$ ns, the difference between the simulated trajectory and the
corresponding energy-density contour is negligible. The scalar product between
these two curves is plotted in Fig. \ref{fig:fig7}(b) as a function of
$m_{z}$, which is not less than 99\%.
These results strongly support the validity of our analysis.

\newpage

%

\end{document}